\documentclass{PoS}
\usepackage{cite}
\usepackage{amsmath}
\usepackage{amssymb}
\usepackage{revsymb}
\usepackage{url}

\title{Curvature of the pseudocritical line in (2+1)-flavor QCD with HISQ fermions}

\ShortTitle{Curvature of the pseudocritical line in QCD}

\author{\speaker{Leonardo Cosmai}%
\\
        INFN - Sezione di Bari, I-70126 Bari, Italy\\
        E-mail: \email{leonardo.cosmai@ba.infn.it}}
\author{Paolo Cea\\
            Dipartimento di Fisica dell'Universit\`a di Bari, I-70126 Bari, Italy and INFN, Sezione di Bari, I-70126 Bari, Italy\\
            E-mail: \email{paolo.cea@ba.infn.it}}
\author{Alessandro  Papa\\
            Dipartimento di Fisica, Universit\`a della Calabria, \\
            \& INFN - Gruppo Collegato di Cosenza, I-87036 Rende, Italy \\
            E-mail: \email{papa@cs.infn.it}}

\abstract{We study QCD with (2+1)-HISQ fermions at nonzero temperature 
and nonzero imaginary baryon chemical potential. 
Monte Carlo simulations are performed using the MILC code 
along the line of constant physics with a light to strange mass ratio of $m_l/m_s=1/20$
on lattices up to $48^3 \times 12$ to check for finite cutoff effects.
We determine the curvature of the pseudocritical line extrapolated to the continuum limit.}

\FullConference{The 33rd International Symposium on Lattice Field Theory\\
		14 -18 July 2015\\
		Kobe International Conference Center, Kobe, Japan*}

\begin{document}

\section{Introduction}
\label{introd}

Quantum ChromoDynamics (QCD) is widely accepted as the theory of strong 
interactions and, as such, must encode all the information needed to
precisely draw the phase diagram 
in the temperature ($T$) - barion chemical potential $(\mu_B)$-plane. 
As a matter of fact, only some corners of it can be accessed by first-principle applications
of QCD, in the perturbative or in the nonperturbative regime. 
The region
of the phase diagram where $\mu_B/(3 T)\lesssim 1$ is within the reach of the
lattice approach of QCD  and one can therefore address, at least inside this region, the 
problem of determining the shape taken by the QCD pseudocritical line 
separating the hadronic from the deconfined phase.
There is no {\it a priori} argument for the coincidence of the QCD 
pseudocritical line with the chemical freeze-out curve: if the deconfined 
phase is realized in the fireball, in cooling down the system first 
re-hadronizes, then reaches the chemical freeze-out. This implies that the 
freeze-out curve lies below the pseudocritical line in the $\mu_B$-$T$ plane. 
It is a common working hypothesis that the delay between chemical freeze-out and 
rehadronization is so short that the two curves lie close to each other and 
can therefore be compared.
Under the assumptions of charge-conjugation invariance at $\mu_B=0$ and 
analyticity around this point, the QCD pseudocritical line, as 
well as the freeze-out curve, can be parameterized, at low baryon densities, 
by a lowest-order Taylor expansion in the baryon chemical potential, as
\begin{equation}
\frac{T(\mu_B)}{T_c(0)}=1-\kappa \left(\frac{\mu_B}{T(\mu_B)}\right)^2\;,
\label{curv}
\end{equation}
where $T_c(0)$ and $\kappa$ are, respectively, the pseudocritical temperature 
and the curvature at vanishing baryon density.

As is well known direct Monte Carlo simulations of lattice QCD at nonzero baryon density
are hindered by the well known ``sign problem'': $S_E$ becomes complex and 
the Boltzmann weight loses its sense. Several ways out of this problem
have been devised (see Ref.~\cite{Philipsen:2005mj,Schmidt:2006us,deForcrand:2010ys,Aarts:2013bla} for a review).
In the present work we use the approach of analytic continuation from imaginary chemical potential.
Our aim is to determine the continuum limit of the curvature $\kappa$ of the pseudocritical line of QCD with nf =2+1 staggered fermions 
at nonzero temperature and quark density. More details are reported in Ref.~\cite{Cea:2015cya}.

The state-of-the-art of lattice determinations of the curvature $\kappa$,
up to the very recent papers of Ref.~\cite{Bonati:2015bha,Bellwied:2015rza}, 
is summarized in Fig.~10 of Ref.~\cite{Bonati:2014rfa}: depending on the
lattice setup and on the observable used to probe the transition, the value 
of $\kappa$ can change even by almost a factor of three. The lattice 
setup dependence stems from the kind of adopted discretization, the lattice 
size, the choice of quark masses and chemical potentials, the procedure
to circumvent the sign problem.
On the side of the determinations of the freeze-out curve, two recent 
determinations~\cite{Cleymans:2005xv,Becattini:2012xb} of $\kappa$, both
based on the thermal-statistical model, but the latter of them including the
effect of inelastic collisions after freeze-out, give two quite different
values of $\kappa$, each seeming to prefer a different subset of lattice
results (see Fig.~3 of Ref.~\cite{Cea:2014xva} for a snapshot of the
situation). 

\section{Numerical results}

We perform simulations of lattice QCD with 2+1 flavors of rooted staggered 
quarks at imaginary quark chemical potential.
We have made use of the HISQ/tree action~\cite{Follana:2006rc,Bazavov:2009bb,Bazavov:2010ru} 
as implemented in the publicly available MILC code~\cite{MILC}, 
which has been suitably modified by us in order to introduce an imaginary quark
chemical potential $\mu = \mu_B/3$. 
In the present study we put 
$\mu=\mu_l=\mu_s$, with $\mu_l$ the light quark chemical potential and
$\mu_s$ the strange quark chemical potential. This means that the Euclidean partition function of the
discretized theory reads
\begin{equation}
\label{Z}
Z=\int [DU] e^{-S_{\rm gauge}} \prod_{q=u,d,s} {\rm det}(D_q[U,\mu])^{1/4} \;,
\end{equation}
where $S_{\rm gauge}$ is the Symanzik-improved gauge 
action and $D_q[U,\mu]$ is the staggered Dirac operator, modified  
for the inclusion of the imaginary quark chemical
(see Ref.~\cite{Bazavov:2009bb} and appendix~A of Ref.~\cite{Bazavov:2010ru}    for the precise definition of
the gauge action and the covariant derivative for highly improved staggered fermions). 

We have simulated the theory at finite temperature, and for several values of 
the imaginary quark chemical potential, near the transition temperature, 
adopting lattices of size $16^3\times 6$, $24^3\times 6$, $32^3 \times 8$, 
$40^3 \times 10$ and $48^3 \times 12$.  
All simulations make use of the rational hybrid Monte Carlo (RHMC) 
algorithm. The length of each RHMC trajectory has been set to  
$1.0$ in molecular dynamics time units.
We have discarded typically not less than one thousand trajectories for each 
run and have collected from {4k to 8k} trajectories for measurements.

The pseudocritical point $\beta_c(\mu^2)$ has been determined as the value 
for which the renormalized disconnected susceptibility of the light quark chiral condensate
divided by $T^2$ exhibits a peak. \\
The bare disconnected susceptibility is given by:
\begin{equation}
\label{chi_dis}
\chi_{l, \rm disc} =
{{n_f^2} \over 16 L_s^3 L_t}\left\{
\langle\bigl( {\rm Tr} D_q^{-1}\bigr)^2  \rangle -
\langle {\rm Tr} D_q^{-1}\rangle^2 \right\}\;,
\end{equation}
 Here $n_f=2$ is the number of light flavors
and $L_s$ denotes the lattice size in the space direction.  The renormalized chiral susceptibility
is defined as: 
\begin{equation}
\label{chi_ren}
\chi_{l, \rm ren} =   \frac{1}{Z_m^2} \;  \chi_{l, \rm disc} .
\end{equation}
The multiplicative renormalization factor $Z_m$  can be deduced
from an analysis of the line of constant physics for the light quark masses.
More precisely, we have~\cite{Bazavov:2010ru}:
\begin{equation}
\label{zeta}
Z_m(\beta) \;  =  \;  \frac{m_l(\beta)}{m_l(\beta^*)}  \; ,
\end{equation}
where the  renormalization point $\beta^*$ is chosen such that:
\begin{equation}
\label{beta*}
\frac{r_1}{a(\beta^*)} \; =  \;  2.37 \; ,
\end{equation}
where the function  $a(\beta)$ is discussed below.

\begin{figure}[tb]
\centering
\includegraphics*[width=0.6\columnwidth]
{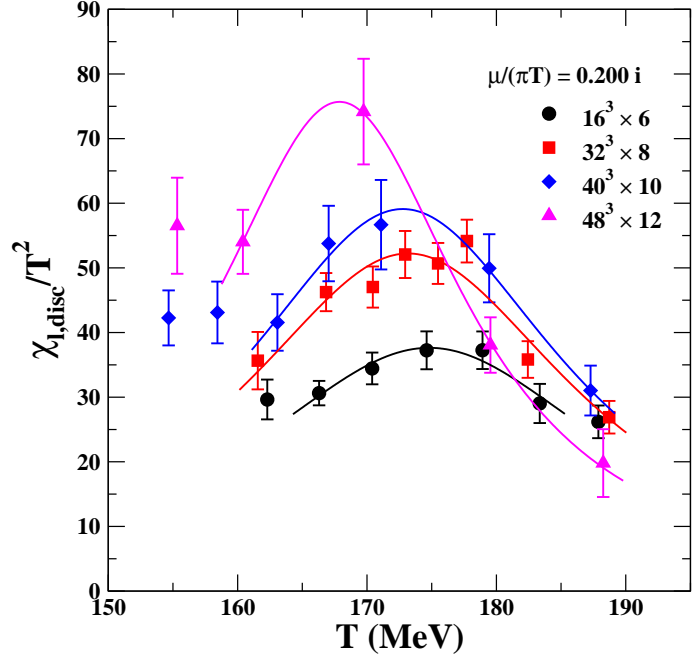}
\caption{The real part of the renormalized susceptibility of the light quark 
chiral condensate over $T^2$ on the lattices $16^3\times 6$,
$32^3\times 8$, $40^3\times 10$ and $48^3\times 12$ at $\mu/(\pi T)=0.2i$. 
Full lines give the Lorentzian fits near the peaks. }
\label{fig_chiral_light_suscep}
\end{figure}

To precisely  localize the peak in $\chi_{l, \rm ren}/T^2$, a Lorentzian fit has been used. For illustrative purposes, 
in Fig.~\ref{fig_chiral_light_suscep} we display our determination of the 
pseudocritical couplings at $\mu/(\pi T)=0.2i$ for all lattices
considered in this work.

To get the ratios $T_c(\mu)/T_c(0)$, we fix the lattice spacing through the 
observables $r_1$ and $f_K$, following the discussion in the Appendix~B of 
Ref.~\cite{Bazavov:2011nk}. \\
 For the  $r_1$ scale  the lattice spacing
is given in terms of the $r_1$ parameter as:
\begin{equation}
\label{scale-r1}
\frac{a}{r_1}(\beta)_{m_l=0.05m_s}=
\frac{c_0 f(\beta)+c_2 (10/\beta) f^3(\beta)}{
1+d_2 (10/\beta) f^2(\beta)} \; ,
\end{equation}
with $c_0=44.06$, $c_2=272102$, $d_2=4281$, $r_1=0.3106(20)\ {\text{fm}}$. \\
On the other hand, in the case of  the  $f_K$ scale we have:
 \begin{equation}
\label{scale-fk}
 a f_K(\beta)_{m_l=0.05m_s}=
\frac{c_0^K f(\beta)+c_2^K (10/\beta) f^3(\beta)}{
1+d_2^K (10/\beta) f^2(\beta)} \; ,
\end{equation}
with $c_0^K=7.66$, $c_2^K=32911$, $d_2^K=2388$, $r_1f_K \simeq 0.1738$. 
In Eqs.~(\ref{scale-r1}) and (\ref{scale-fk}), $f(\beta)$ is the two-loop beta 
function,
\begin{equation}
\label{beta function}
f(\beta)=(b_0 (10/\beta))^{-b_1/(2 b_0^2)} \exp(-\beta/(20 b_0))\;,
\end{equation}
$b_0$ and $b_1$ being its universal coefficients. 

For all lattice sizes 
the behavior of $T_c(\mu)/T_c(0)$ can be nicely fitted with a linear function 
in $\mu^2$,
\begin{equation}
\label{linearfit}
\frac{T_c(\mu)}{T_c(0)} = 1 + R_q \left(\frac{i \mu}{\pi T_c(\mu)}\right)^2 \;,
\end{equation}
which gives us access to the curvature $R_q$ and, hence, to the curvature
parameter $\kappa=-R_q/(9\pi^2)$ introduced in Eq.~(\ref{curv}). 
On the $24^3\times6$ lattice the linearity in $\mu^2$ has been assumed to hold, 
in order to extract $R_q$ from the only available determination at 
$\mu/(\pi T)=0.2i$.

\begin{figure}[tb]
\centering
\includegraphics*[width=0.6\columnwidth]
{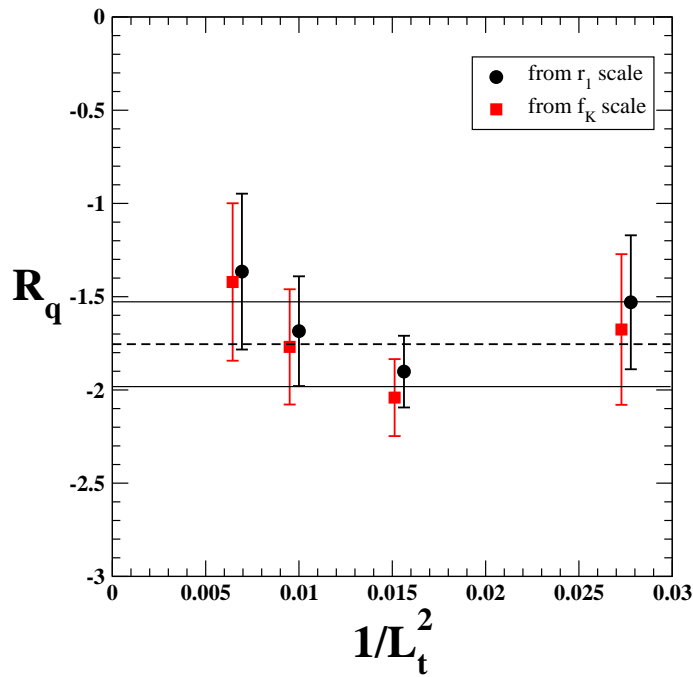}
\caption{Determinations of the curvature $R_q$ on the lattices $24^3\times 6$, 
$32^3 \times 8$, $40^3 \times 10$, $48^3 \times 12$, and from the two different 
methods to set the scale, {\it versus} $1/L_t^2$. The dashed horizontal line
 gives the result of the fit to all data with a constant; the solid 
horizontal lines indicate the uncertainty on this constant.}
\label{curvature_vs_Nt}
\end{figure}

For the sake of the extrapolation to the continuum limit, in 
Fig.~\ref{curvature_vs_Nt} we report our determinations of $R_q$ on the 
lattices $24^3\times 6$, $32^3\times 8$, $40^3\times 10$, $48^3\times 12$,  
and from the two different methods to set the scale, {\it versus} $1/L_t^2$.

Within our accuracy, cutoff effects on $R_q$ are negligible, so that a 
constant fit works well over the whole region ($\chi^2_r \simeq 0.99$), thus 
including also the smallest $24^3\times6$ lattice. Taking into account
the uncertainties due to the continuum limit extrapolation, 
\begin{equation}
\label{curvature}
\kappa = 0.020(4) \;.
\end{equation}
Therefore we can conclude that, within the accuracy of our determinations, cutoff effects
on the curvature are negligible already on the lattice with temporal size
$L_t=6$. Our determination of the curvature parameter, $\kappa$=0.020(4),
is indeed compatible with the value quoted in our previous 
paper~\cite{Cea:2014xva}, $\kappa$=0.018(4), without the extrapolation to
the continuum.

\begin{figure}[tb]
\centering
\includegraphics*[width=0.6\columnwidth]
{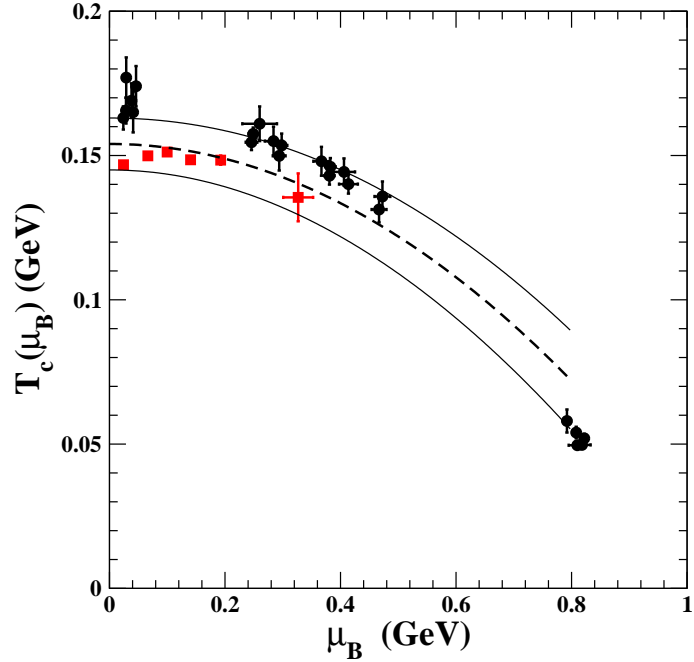}
\caption{$T_c(\mu_B)$ versus $\mu_B$ (units in GeV). 
Experimental values of $T_c(\mu_B)$ are taken
from Fig.~1 of Ref.~\cite{Cleymans:2005xv} (black circles) and from
Fig.~3 of Ref.~\cite{Alba:2014eba} (green triangles),  for the 
standard  hadronization model and for the  susceptibilities of conserved charges respectively. 
The dashed line is a parametrization  corresponding to
$T_c(\mu_B) = T_c(0) - b \mu_B^2$ with $T_c(0)=0.154(9)\, \text{GeV}$ and 
$b=0.128(25)\,{\text{GeV}}^{-1}$ . 
The solid lines represent the corresponding  error band.}
\label{Tcmu}
\end{figure}

Finally it is interesting to extrapolate the critical line as determined in this work 
to the region of real baryon density and compare it with the freeze-out
curves resulting from a few phenomenological analyses of relativistic
heavy-ion collisions. This is done in Fig.~\ref{Tcmu}, where we report
two different estimates. The first is from the analysis of 
Ref.~\cite{Cleymans:2005xv}, based on the standard statistical hadronization 
model, where the freeze-out curve is parametrized as 
\begin{equation}
\label{cleymans}
T_c(\mu_B) = a - b \mu_B^2 - c \mu_B^4\;,
\end{equation}
with $a=0.166(2) \ {\text{GeV}}$, $b=0.139(16) \ {\text{GeV}}^{-1}$, 
and $c=0.053(21) \ {\text{GeV}}^{-3}$.
The second estimate is from Ref.~\cite{Alba:2014eba} and is based on the analysis
of susceptibilities of the (conserved) baryon and electric charges.
In fact, our critical line is in nice agreement with all the freeze-out points of
Refs.~\cite{Cleymans:2005xv,Alba:2014eba}. In particular, using our estimate
of the curvature,  Eq.~(\ref{curvature}), we get   
$b=0.128(25) \ {\text{GeV}}^{-1}$, in very good agreement with the
quoted phenomenological  value.

Some {\it caveats} are in order here. We do not expect our critical line 
to be reliable too far from $\mu=0$: as a rule of thumb, we can trust it up 
to real quark chemical potentials of the same order of the modulus
of the largest imaginary chemical potential included in the 
fit~(\ref{linearfit}), {\it i.e.} $|\mu|/(\pi T)=0.25$. This translates
to real baryon chemical potentials in the region $\mu_B/T\lesssim 0.25$.
Moreover, the effect of taking $\mu_s=\mu_l$ instead of $\mu_s < \mu_l$
should become visible on the shape of the critical line as we move away
from $\mu=0$ in the region of real baryon densities, thus reducing
further the region of reliability of our critical line. So, from a prudential
point of view, the agreement shown in Fig.~\ref{Tcmu} could be considered
the fortunate combination of different kinds of systematic effects. We cannot
however exclude the possibility that the message from Fig.~\ref{Tcmu} is to be 
interpreted in positive sense, {\it i.e.} the setup we adopted and the
observable we considered may catch better some features of the crossover
transition, thus explaining the nice comparison with freeze-out data.
Indeed,  our  result for the  continuum extrapolation of the curvature $\kappa$ is
in fair agreements with the recent estimates in  Ref.~\cite{Bonati:2015bha}, where
both setup  $\mu_s=\mu_l$ and $\mu_s=0$ were adopted,
and Ref.~\cite{ Bellwied:2015rza},  where the strangeness neutral trajectories
were determined from lattice simulations  by imposing $\langle n_S \rangle=0$.

\section*{Acknowledgements}

This work was based in part on the MILC Collaboration's public lattice gauge theory code ({\url{http://physics.utah.edu/~detar/milc.html}) and has been partially supported by INFN SUMA project.
Simulations have been performed on BlueGene/Q at CINECA (IscraB\_EXQCD and CINECA-INFN agreement).

\bibliography{hd_qcd}
\providecommand{\href}[2]{#2}\begingroup\raggedright\endgroup

\end{document}